\documentclass[]{spie}  

\usepackage{amsmath,amsfonts,amssymb}
\usepackage{graphicx}
\usepackage[colorlinks=true, allcolors=blue]{hyperref}

\usepackage{siunitx}  
\usepackage{upgreek} 
\usepackage{lettrine}

\newcommand{\ket}[1]{\left| #1\right\rangle}

\title{Pulse, polarization and topology shaping of polariton fluids}

\author[a]{Lorenzo Dominici}
\author[b]{David Colas}
\author[b]{Stefano Donati}
\author[c]{Galbadrakh Dagvadorj}
\author[a]{\\Antonio Gianfrate}
\author[b]{Carlos S\'{a}nchez Mu\~{n}oz}
\author[a]{Dario Ballarini}
\author[a]{Milena De Giorgi}
\author[a]{\\Giuseppe Gigli}
\author[d]{Marzena H.~Szyma\'{n}ska}
\author[b]{Fabrice P.~Laussy}
\author[a]{Daniele Sanvitto}

\affil[a]{CNR NANOTEC, Istituto di Nanotecnologia, Via Monteroni, 73100 Lecce, Italy}
\affil[b]{Departamento de F\'{i}sica Te\'{o}rica de la Materia Condensada, UAM, 28049 Madrid, Spain}
\affil[c]{Department of Physics, University of Warwick, CV47AL Coventry, UK}
\affil[d]{Department of Physics and Astronomy, University College London, WC1E6BT London, UK}

\authorinfo{L.D.: lorenzo.dominici@gmail.com, F.P.L.: fabrice.laussy@gmail.com ---
This paper was originally published in:\\ Complex Light and Optical Forces XI, edited by David L.~Andrews, Enrique J.~Galvez, Jesper Gl\"{u}ckstad, \\Proc.~of SPIE Vol.~10120, 101200E \copyright\ 2017 SPIE CCC code: 0277-786X/17/\$18 · doi: 10.1117/12.2250997. \\ One print or electronic copy may be made for personal use only. Systematic or multiple   reproduction, distribution to multiple locations via electronic or other means, duplication   of any material in this paper for a fee or for commercial purposes, or modification of the content of the paper are prohibited.
The final published version is available at: 
  \url{https://doi.org/10.1117/12.2250997}}

\begin{document} 
\maketitle

\begin{abstract}
{\large \textbf{
Here we present different approaches to ultrafast pulse and polarization shaping, based on a  ``quantum fluid'' platform of polaritons.
Indeed we exploit the normal modes of two dimensional polariton fluids made of strong coupled quantum well excitons and microcavity photons,
by rooting different polarization and topological states into their sub-picosecond Rabi oscillations.
Coherent control of two resonant excitation pulses allows us to prepare the desired state of the polariton,
taking benefit from its four-component features given by the combination of the two normal modes with the two degrees of polarization. 
An ultrafast imaging based on the digital off-axis holography technique is implemented to study the polariton complex wavefunction with time and space resolution. 
We show in order coherent control of the polariton state on the Bloch sphere, an ultrafast polarization sweeping of the Poincar\'{e} sphere, and the dynamical twist of full Poincar\'{e} states such as the skyrmion on the sphere itself.
Finally, we realize a new kind of ultrafast swirling vortices by adding the angular momentum degree of freedom to the two-pulse scheme.
These oscillating topology states are characterized by one or more inner phase singularities tubes which spirals around the axis of propagation.
The mechanism is devised in the splitting of the vortex into the upper and lower polaritons, resulting in an oscillatory exchange of energy and angular momentum 
and in the emitted time and space structured photonic packets.}}

\end{abstract}


\section{INTRODUCTION}
\label{sec:intro}  

\lettrine{W}{e}
make use of microcavity polaritons~\cite{Byrnes2014}, 
a ``quantum fluid'' platform 
constituted by bosonic hybrid quasiparticles
of strongly coupled excitons and photons fields,
in order to achieve different kind of intensity, polarization and topology shaping.
Due to the photonic outcoupling of the polaritons, 
which makes possible their resonant excitation and detection,
such time and space structured-field features are tranferred to the emitted optical pulses too.
The realization of exciton polariton condensates
in semiconductor microcavities~\cite{Kasprzak2006,Balili2007} has
paved the way for a prolific series of studies into quantum
hydrodynamics in two-dimensional
systems~\cite{whittaker_polariton_2016,dominici_real-space_2015,Dreismann2014,Roumpos2012,amo_polariton_2011,Pigeon2011,sanvitto_persistent_2010,amo_superfluidity_2009,lagoudakis_quantized_2008}.
Microcavity polaritons are intriguing systems for the
study of topological excitations in nonequilibrium interacting superfluids,
also thanks to  
rich spinorial patterns
and polarization splitting terms leading to typical spin-orbit coupling~\cite{Manni2011,Toledo2014,Flayac2010a}. 
In very recent works~\cite{dominici2014,colas_polarization_2015}, thanks to significant progress in both the quality of
the structures and in the laboratory state of the art, we have been able to
both observe and control the microcavity polariton Rabi dynamics.
We can span from Rabi
oscillating configurations to eigenstate superpositions, and control
them by multiple optical pulses that can amplify or switch states. 
Moreover, by exploting the spin degree of freedom,
we have been able to tranfer the intensity oscillations into polarization oscillations, 
thus making such systems even more suitable for polarization shaping
engineering and applications.
Space structured topology such as full Poincar\'{e} skyrmion beams have be imprinted too 
and their reshaping described in the polarization space~\cite{donati_twist_2016}.
The most recent of our experiments and models are hence pushed further on,
by empowering with the Rabi oscillations the ultrafast topology shaping of quantized vortices,
as shown in the last section of this work.

  \begin{figure} [ht]
   \begin{center}
   \begin{tabular}{c} 
   \includegraphics[height=9cm]{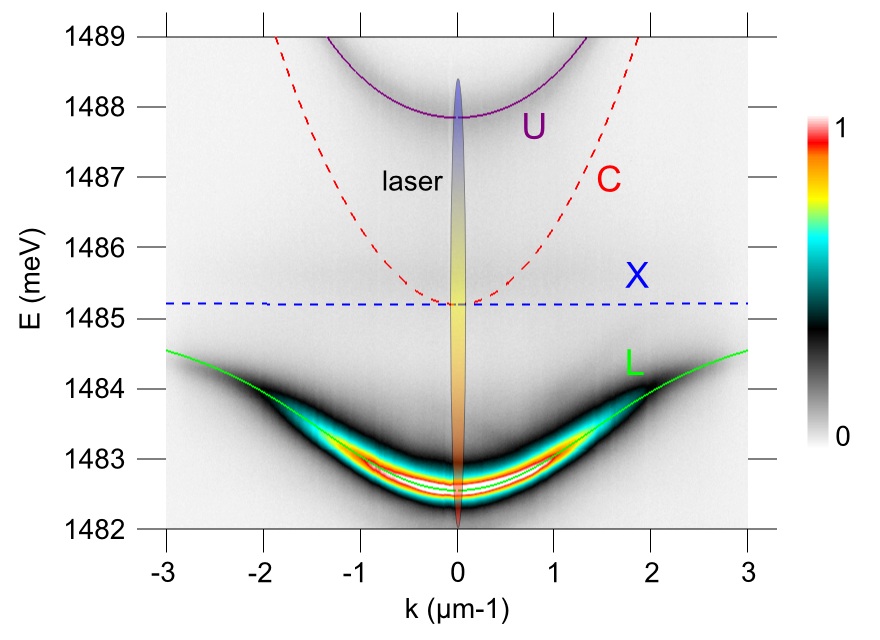}
	\end{tabular}
	\end{center}
   \caption[example] 
   { \label{fig:polariton_modes} 
Experimental polariton dispersion. The emission is obtained after off-resonant excitation. The bare modes are denoted as dashed lines, microcavity mode (C, red) and quantum well exciton (X, blue). The polariton normal modes are denoted as solid lines, UPB (U, violet) and LPB (L, green). A sketch of a resonant femtosecond laser tuned to excite both LPB and UPB branches is shown as a gradient color ellipse at $k = 0$. Data readapted from Dominici et al.~(2015)~\cite{dominici_real-space_2015}.}
   \end{figure}

\section{POLARITON RABI OSCILLATIONS AND COHERENT CONTROL}

\lettrine{A}{n} important peculiarity of polaritons here is at play,
the so called Rabi oscillations~\cite{dominici2014,colas_polarization_2015}.
Indeed, a strong coupling of the two bare oscillators, 
the microcavity photons (C) and the quantum well exciton (X) fields,
results also in their splitting into two normal modes of the system.
These are the upper (UPB) and lower polariton branch (LPB) modes, 
which differ in energy and dephasing time (which is, mode bandwidth),
and where the two bare fields oscillate with in-phase and phase-opposition, respectively. 
The simultaneous excitation of the two modes
can be perceived as a time beating in their projection upon the photon and exciton basis.
A complementary point of view is the continuous transformation of photons into excitons and then back.
Rabi oscillations have been known since the early pioneering works on polaritons,
but their space resolved ultrafast imaging only recently achieved,
and never before exploting the high degree of coherence as here.
Typical bare and normal modes of microcavity polaritons are shown in Fig.~\ref{fig:polariton_modes}.

Cyclical exchange of excitation between two coupled modes ~$a$ and~$b$
can occur at either the single particle or many particle levels.
When this occurs at the single particle level
in a quantum two-level system, it provides the ground for a
qubit~\cite{schumacher95a}, which, if it can be further manipulated,
opens the possibility to perform quantum information
processing~\cite{nielsen_book00a}.  Such an oscillation is of
probability amplitudes and therefore is a strongly quantum mechanical
phenomenon, that involves maximally entangled states:
\begin{equation}
  \label{eq:lunjul21152706CEST2014}
  \ket{\Psi(t)}=\alpha(t)\ket{1_a,0_b}+\beta(t)\ket{0_a,1_b}\,.
\end{equation}
The same physics also holds, not at the quantum level, but with
coherent states of the fields, a situation known in the literature as
implementing an ``optical atom''~\cite{spreeuw93a} or a ``classical
two-level system''~\cite{faust13a}.  The oscillation is then more
properly qualified as ``normal mode
coupling''~\cite{zhu90a,khitrova06a} as it is now between the fields
themselves:
\begin{equation}
  \label{eq:marjul1094910CEST2014}
  \ket{\psi(t)}=\ket{\alpha(t)}\ket{\beta(t)}\,,
\end{equation}
rather than their probability amplitudes.  The terminology of Rabi
oscillations is however used also in the classical
case~\cite{matthews99b,vasa13a}, such for example
the purpose to realize classical bits, 
``cebit''~\cite{spreeuw98a}. 
or other classical counterparts of the quantum
version~\cite{dragoman_book04a}. 
As a futher example, classical two-level systems~\cite{spreeuw90a} 
were recently pursued in the rising field of nanomechanical
optics~\cite{faust13a,okamoto13a}. 
In our case, the polariton system~\cite{kavokin_book11a} 
provides hence a good platform to implement two-level dynamics at both the
quantum~\cite{hennessy07a} and classical level~\cite{weisbuch92a}. 
As said, the polariton nature itself is that of a two-level system, 
giving the strong coupling regime between the cavity photon and a semiconductor
exciton, also named bare oscillators. 
In the specific, two-dimensional microcavities 
embedding inorganic quantum wells (QWs)~\cite{carusotto13a}
have attracted attention for their ability 
to reach the regime of
Bose-Einstein condensation~\cite{kasprzak06a} and
superfluidity~\cite{amo_superfluidity_2009,amo_collective_2009},
with a plethora of collective wavefunction phenomena
hydrodynamics including coherence, quantized vorticity, strong nonlinearities
and real-space pattern formation~\cite{liew08a,amo_exciton-polariton_2010,amo_polariton_2011,dominici_real-space_2015,whittaker_polariton_2016}. 
On the application side, the most intriguing
perspectives involve schemes and demonstrations for new polariton lasers~\cite{baumberg08a,schneider13a}
and all-optical transistors and logical operations~\cite{ballarini_all-optical_2013}. 
Pioneering attempts to observe the Rabi
oscillations at the core of the polariton physics, 
encountered intrisic difficulties represented, e.g., by their sub-ps time range 
and reported very few
oscillations with orders of magnitude visibility loss each cycle~\cite{norris94a}.
The inhomogeneous broadening of
excitons~\cite{savona96b}, not taking into account the space dephasing
of the space-integrated signals, could provide a
qualitative agreement only. Later on, nonlinear pump-probe
techniques~\cite{wang95b,marie99a,huynh02b,brunetti06a}, achieved a better visibility in the registration
of the modes beating but didn't provided the necessary degree for a deeper analysis. 
Our recent experiments could image the time and spatially resolved oscillations and achieve Rabi piloting 
by two pulse schemes able to achieve the
wished modes superposition including a polarization degree of freedom.
Such schemes are of paramount importance~\cite{ridolfo11a} 
also in light of the possibility to address polaritons at the single
particle level~\cite{boulier_polariton-generated_2014,cuevas_entangling_2016}.

 \begin{figure} [ht]
   \begin{center}
   \begin{tabular}{c} 
   \includegraphics[height=11cm]{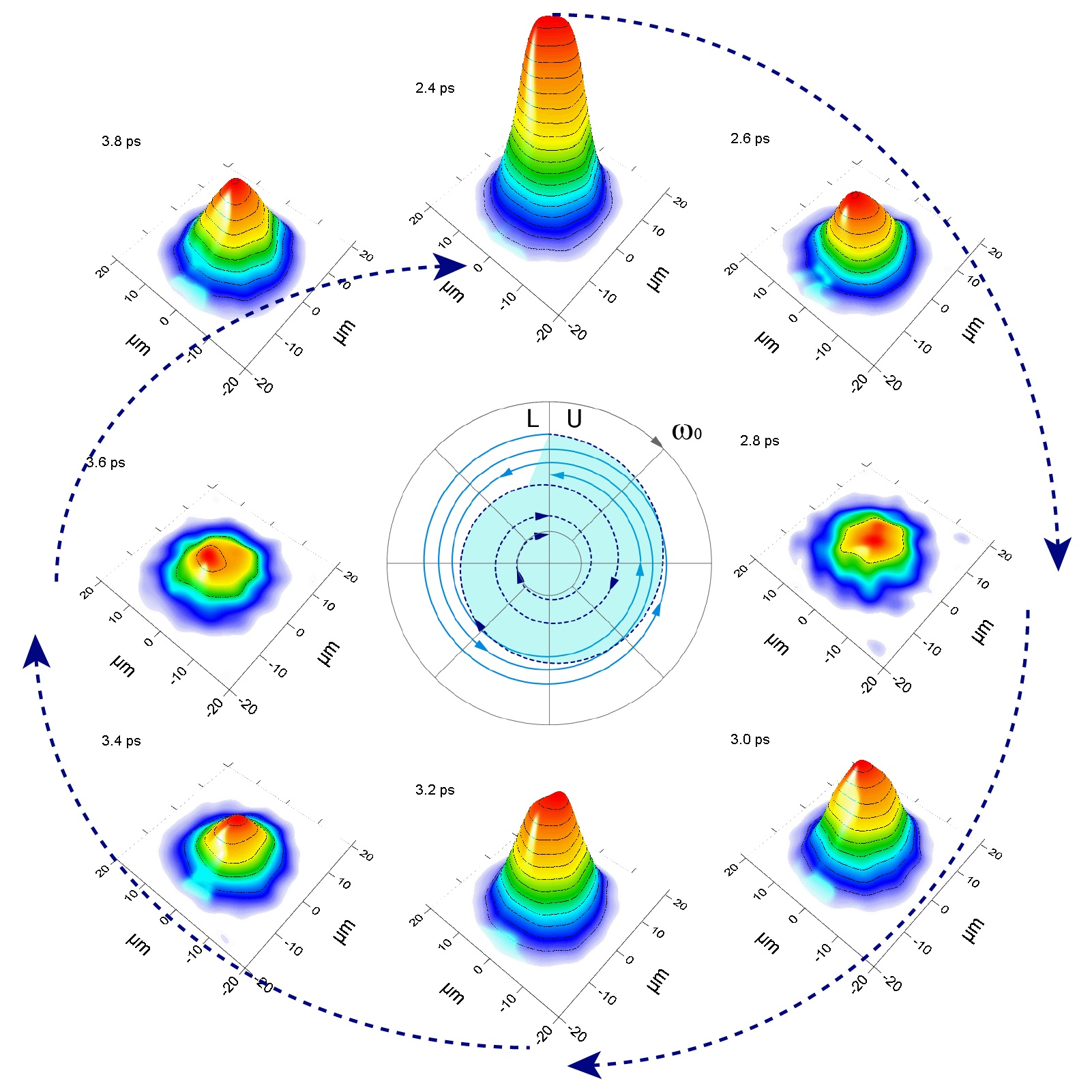}
   \end{tabular}
   \end{center}
   \caption[example] 
   { \label{fig:spacetime_Rabi} 
Two Rabi cycles seen through the time and space resolved photonic emission (outer panels) and the scheme of the UPB and LPB modes depiceted in the $\omega_0$ rotating frame of reference (inner panel). The time reference is given by the central frequency between the bare modes ($\omega_0 = \frac{\omega_C + \omega_X}{2}$).}
   \end{figure}

\subsection{Single pulse Rabi oscillations}

As shown in Fig.~\ref{fig:polariton_modes},
the $130$~fs long and $8$~nm energy broad
laser pulse spreads in energy over the two branches, preparing the system in
a bare state and not in an eigenstate, inducing oscillations between its two components.
One can see on the outer panels of Fig.~\ref{fig:spacetime_Rabi} a typical experimental observation
of the space resolved photonic emission from the microcavity sample.
As expected,
the photon field arises and vanishes cyclically, during each Rabi cycle.
The different dephasing times of the two modes, suggested by the inner scheme of the figure
(with different radius of the rotating modes), is leading also to a damping of the oscillations between consecutive cycles.
Here the polaritons have quite a large spatial extension, given by the
exciting laser, $\approx 10 \upmu m$. 
However, given the system is linear, presents a minimal diffusion
and is not imparted by any momentum, the dynamics can be reduced
to zero dimension (i.e., the field is homogeneous and with no space reshaping, 
there is no exploitation of the space degree of freedom). 
Indeed, the excitation power was set
at a low enough density in order to maintain their bosonic properties in the linear
regime. 
We could have thus access to the subpicosecond linear Rabi
oscillations through the coherent fraction $|\psi_C(r,t)|^2$ of the photonic field. 

\begin{figure} [ht]
   \begin{center}
   \begin{tabular}{c} 
   \includegraphics[height=6.5cm]{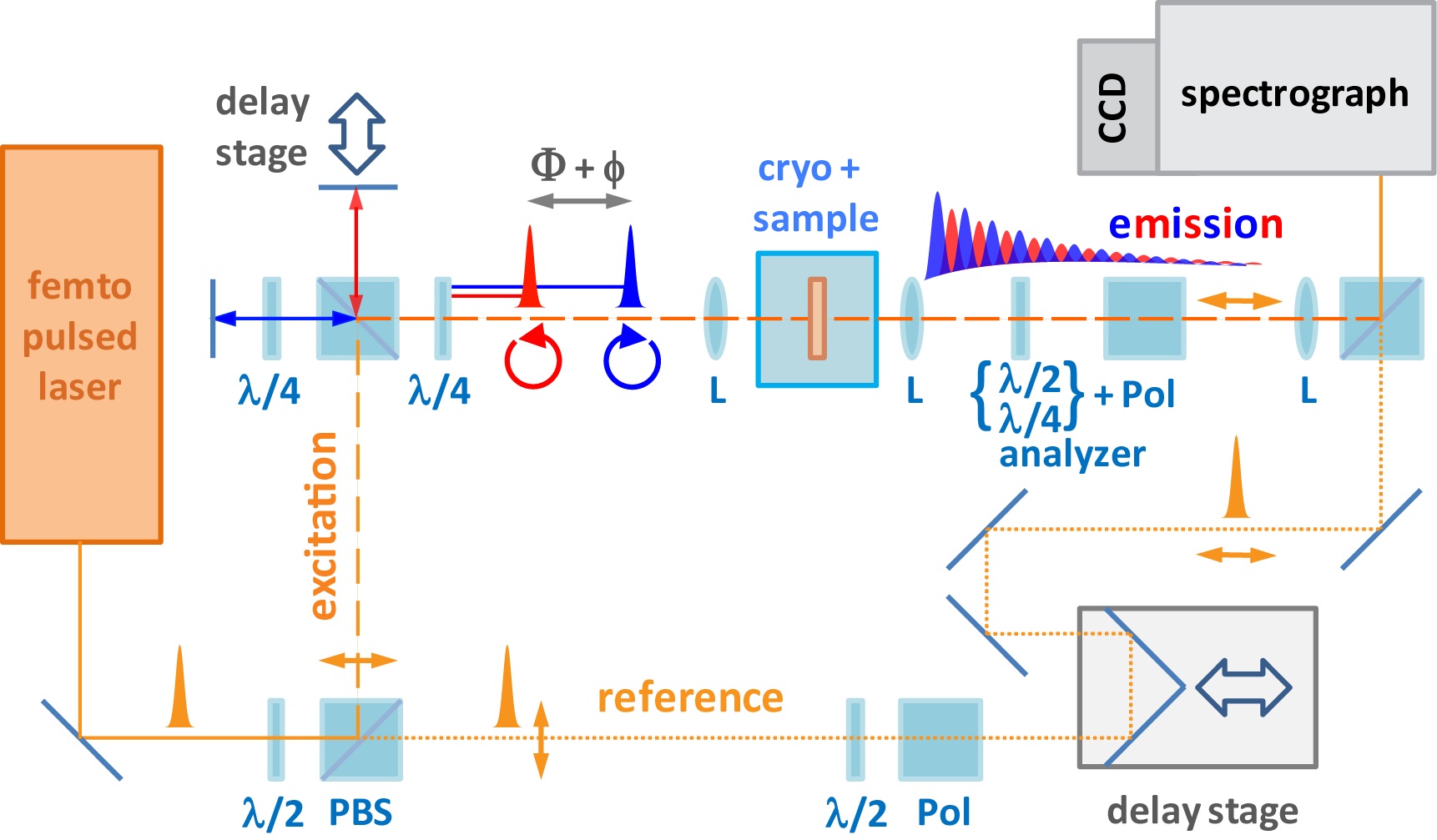}
	\end{tabular}
	\end{center}
   \caption[example] 
   { \label{fig:exp_setup} 
Experimental setup for double pulse coherent control experiments with polarization control and digital off-axis holography imaging.
The interference between the device emission and the reference arms is implemented through a Mach-Zehnder configuration, while the two-pulse excitation is achieved in a Michelson configuration.}
   \end{figure}

To this extent, our setup can access the complex wavefunction, 
i.e., measuring both its amplitude and phase, by holography, a
technique of increasing use to image polariton fluids for which both
of these informations are of crucial
importance~\cite{nardin11a,anton12a}. We recourse to a variation known
as off-axis digital holography~\cite{schnars_book05a}, which provides
high-quality results by separation of the diffracted images of an
off-axis reference frame and the signal. We adapted it to support
ultrafast and tunable multiple-pulse experiments, as shown in Fig.~\ref{fig:exp_setup}.
As such, our measurement does not rely on nonlinear detection scheme or interactions, as
in previous works~\cite{wang95b,marie99a,huynh02b}, but on
interferences only. 
The photonic field that is accessible experimentally, 
 have been fitted with the
optical model developed in detailed in~\cite{dominici2014}, 
and its excitonic
counterpart recovered through the polaritonic equations .

\subsection{Two-pulse coherent control}

\begin{figure} [ht]
   \begin{center}
   \begin{tabular}{c} 
   \includegraphics[height=8.5cm]{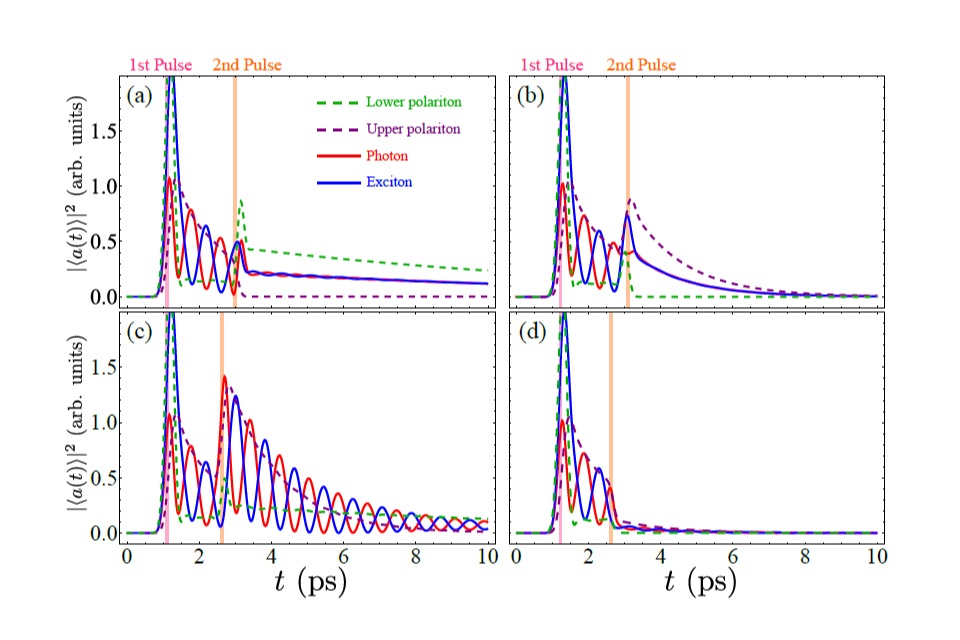}
	\end{tabular}
	\end{center}
   \caption[example] 
   { \label{fig:coherent_control} 
Two-pulse experiment as seen through all the theoretical variables: bare states in solid
and eigenstates in dashed lines. The experimentally available variable is the photon field, in red.
The four cases correspond to: (a) switching-off of the Rabi oscillation by bringing the state into a
LPB condensate, (b) switching-off of the Rabi oscillation by bringing the state into a
UPB condensate, (c) enhancement of the oscillations and (d) annihilation of the field. 
Reprinted with permission from Dominici et al.~(2014).~\cite{dominici2014} Copyright (2014) by the American Physical Society.}
   \end{figure}

With such an accurate command of the system, we are able to time
precisely the arrival of a second pulse and perform a comprehensive
coherent control on the coupled dynamics.
The Rabi oscillations are triggered by sending a
first pulse quite broad in energy that initiates a dominant photon or exciton fraction.
One can refine the state by sending a second pulse, similar to the first one, to
any desired configurations.   
The delay of the second photonic pulse can be set
according to four different main preparations,
depending on the combination of the pulses' relative delay 
in both the Rabi period order $\Phi$ and optical period order $\phi$. 
For example, the Rabi antiphase condition is achieved 
upon coupling the second pulse when the cavity field is empty and the state
is fully excitonic. In such case, injecting a second fully photonic pulse in optical
(resp.~anti-optical) phase with the exciton, for instance, creates an
LP (resp.~UP), as shown in Fig.~\ref{fig:coherent_control}(a) (resp.~Fig.~\ref{fig:coherent_control}(b)). 
The transformation of a bare state, created by the 1st pulse, into a LPB or UPB
state when the second pulse arrives, is associated to the switching off of the Rabi oscillations, 
and the population decreases with the lifetime of the respective mode.
In Fig.~\ref{fig:coherent_control}(c),
the second pulse is arriving during the photonic phase of the condensate (Rabi in-phase condition),
and also in optical phase with the cavity field. 
This is a case of Rabi oscillations
enhancement, where the oscillations cycle is magnified with the second pulse.
In the case of Fig.~\ref{fig:coherent_control}(d) instead the second pulse annihilates 
the field.
This is achieved by sending a pulse optically out of phase but in phase with the
Rabi oscillations, inducing destructive interferences that cancel the field intensity.
All these cases demonstrate the possibility to
do coherent control of the strong light-matter coupling dynamics.
Such proof of principle fully demonstrated here also suggests that there
would be no fundamental difficulty in sending more than two pulses. Interesting
perspectives are now opened, notably at the single-particle level to perform quantum
information processing.

\section{RABI POLARIZATION SHAPING}
\label{sec:sections}

\lettrine{W}{e} here discuss how to generate new states of light by using this most fundamental of
properties from the polaritons, their Rabi oscillations.
The realization is based on the proposal to link these two concepts
of polarized light by using the oscillating property of the microcavity polaritons.
The idea is the following: a
light whose polarization is varying in time can be emitted from a microcavity by
exciting it with two orthogonally polarized pulses and with a suitable time delay
between them. Working with orthogonally polarized pulses ensures that no interactions
between the fields of different polarizations will occur. The superposition
of the oscillating fields results in the precession of the Stokes vector of the emitted
light. Polaritons being particles with finite lifetime, it results in a drift of the
polarization from a circle on the Poincar\'{e} sphere (whose radius and position can
be controlled) to a single point at long time. This leads to the generation of a new
kind of polarized light, suitable to cover selected area of the Poincar\'{e} sphere, and
possibly the whole sphere, in a range of a few picoseconds.
The effect has been demonstrated as a proof of principle
with microcavity polaritons at the femtosecond time scale in our previous work.
 It shows the emission of
time varying polarized light, that, for instance, can cover a whole hemisphere of
the Poincar\'{e} sphere. 
 But as it relies on the Rabi oscillations, this effect could be thus transposed and
obtained with other platforms operating at different time scales (from attoseconds to milliseconds) and exploiting the same property.

\subsection{Polarization Shaping}

The first realization of a shaped pulse produced light whose intensity, momentary
frequency and light polarization were varied as a function of time in a timescale of
femtoseconds. This was reported by Brixner and Gerber in 2001~\cite{brixner_femtosecond_2001}. 
Thereby was opened
the era of ''polarization pulse shaping'' with a huge number of potential applications.
Before that,
only simple polarization profiles could be achieved by using interferometric combinations
of two polarization components. It has been demonstrated by Zhuang et
al. (1997)~\cite{zhuang_polarization_1999} that a stack of three homogeneous nematic liquid-crystal cells, or
Liquid-Crystal display (LCD) could be used as a controller to bring any state of
polarization of light from one arbitrary state to another. The idea of Brixner and
Gerber was thus to build a device based on a set of many-pixel (256) two-layer
LCDs to obtained a time varying polarized light, each pixel changing the state of
polarization (SOP). By applying a suitable voltage to the separated pixels, one tune
individual frequency intervals throughout the laser spectrum. The phases differences
leasing to a complex polarization-modulated laser pulses in the time domain.
The desired sequence of voltage modulations, that correspond to the various SOPs,
can be computer-controlled with a corresponding algorithm.

The range of applications is potentially huge since the interaction between light
and matter is polarization sensitive. This technique was thus used to increase the
performances of many applications based on polarization. In a later study, Brixner
et al. (2004)~\cite{brixner_quantum_2004} reported the ionization of potassium dimer molecules, where
dipole transitions are favoured by different directions of the exciting laser field.
With this example, we understand directly the advantages of having a single laser
beam whose polarization varies in time. Polarization pulse shaped lasers are also
an efficient method for nano-optical manipulation. Aeschlimann et al. (2007)~\cite{aeschlimann_adaptive_2007}
have achieved subwavelength dynamic localisation of electromagnetic intensity on
the nanometre scale, overcoming the spatial restrictions of conventional optics by
using adaptive polarization shaping of femtosecond laser pulses. This technique
has also been used for the generation of ultrashort laser pulse pairs, whose time
delay between them is controlled with a zeptosecond precision~\cite{kohler_zeptosecond_2011}. Dozens of
other further studies based on this principle have been done in the last fifteen years,
feeding a rich literature.
This technique of pulse shaping allows one to reach a huge variety of different
time varying polarized pulses, but it suffers also from some limitations. The
number of SOP contained in the resulting pulse depends directly on the number of
LCD, as they are independently tuned to obtain the desired ellipticity. This cannot
be enough to cover entire parts of a Poincar\'{e} sphere. It also requires a complex
setup, involving LCDs, all-reflective zero-dispersion compressor, interferometers
and computer resources, the desired polarization being computed with special algorithms.
Recent advances in polarization state generation (PSG) 
 involve arbitrary polarization synthesis by a plasmonic nanoantenna
fed by the coherent control of two input waveguides~\cite{rodriguez-fortuno_universal_2014}
and parallel architecture based on digital micromirror devices~\cite{she_parallel_2016}.

Another class of polarized beams has been theoretically developed later in
order to cover the whole Poincar\'{e} sphere of polarization. The so-called ''Full
Poincar\'{e} Beams'' have been introduced by Beckley et al. (2010)~\cite{beckley_full_2010} and consists
in a superposition of a Gaussian mode and a spiral-phase Laguerre-Gauss mode
having orthogonal polarizations. In this case, and unlike the Brixner pulses, all the
polarization states are displayed spatially into the beam.

 \begin{figure} [ht]
   \begin{center}
   \begin{tabular}{c} 
   \includegraphics[width=8.5cm]{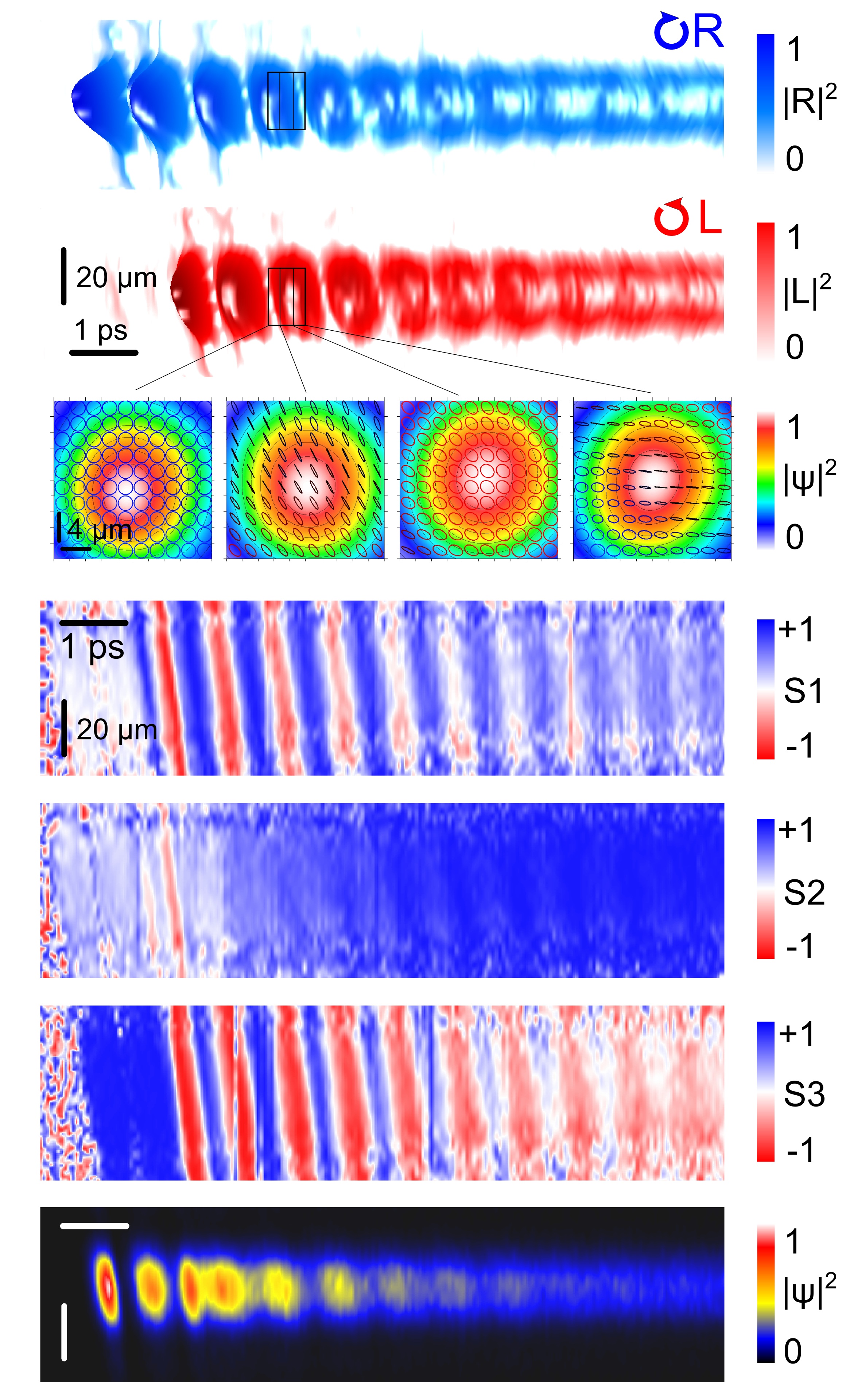}
	\end{tabular}
	\end{center}
   \caption[example] 
   { \label{fig:pol_shaping} 
Experimental polarized Rabi oscillations and contro-polarized two pulse coherent control for polarization shaping. 
The top two panels show the time-space charts of the polariton amplitude for the right ($R$, blue)
and left ($L$, red) circular polarization. Their Rabi oscillations remain decoupled.
In the third row the spatial distributions of the density and polarization texture at
200 fs time intervals during one of the initial cycles after the second pulse arrival. 
The emitted intensity scale in each panel has been normalized to its top density but it is remains basically constant (color scale in
the bottom row chart),  after the second pulse arrival 
On the other hand, the oscillations are transferred to the polarization texture.
The fourth, fifth and sixth panels show the time-space charts for the three Stokes parameter,
$S_1$, $S_2$ and $S_3$. The Rabi oscillations in such panels are clearly seen after the arrival of the second pulse.
The last panel shows the overall density. Data readapted from Colas et al.~(2015)~\cite{colas_polarization_2015}. 
}
   \end{figure}

\subsection{Polariton sweeping on the Poincar\'{e} Sphere}
We can now take advantage of a feature that is usually regarded as a shortcoming
of microcavity polaritons, but that in our case will turn the simple effect just
proposed into a mechanism that powers a new type of light. 
 This effect would
thus provide both of the mechanisms we have mentioned before, namely (i) full
Poincar\'{e} beams, (ii) in time. 
In the panels of Fig.~\ref{fig:pol_shaping} is highlighted the realization
of a contro-polarized two-pulse coherent control experiment.
This is achieved through the full exploitation of the setup previously introduced,
(see the details of the two excitation arms with polarization control in Fig.~\ref{fig:exp_setup}).
The top two rows in Fig.~\ref{fig:pol_shaping} report the time behavior of the cross-section
density of the polariton emission resolved in the two spin populations,
right (R, blue, first panel) and left (L, red, second panel) circular. 
The starting time of the two spin emission is correspondent to the arrival of the two pulses, respectively.
As it can be seen the two spin populations oscillates mutually independent, 
nine or ten oscillations can be resolved before the polariton dephasing leads to the oscillations quench.
Here we have set a relative time delay to achieve the condition of anti-Rabi phase, $\Phi = 1.5~T_R$.
Such half-integer value is a good condition to transfer the density oscillations in each of the two spin
to polarization oscillations.
The resulting polarization texture can be observed in the real-space maps of the third row,
where the normalized total density profile is overlapped too in a central square area of $20~\text{x}~20~\upmu\text{m}^2$.
The four time snapshots, at almost regular intervals of 200 fs, corresponds to the four vertical lines 
of the small boxes in the upper two rows.
We see that right circular, antidiagonal linear, left circular and horizontal linear states are alternating in the ultrafast sequence.
The associated Stokes parameter are evolving in time as highlighted by the following three time-space panels. 
Finally the bottom row, presenting the total density through time,
shows that the initially high visibility Rabi time fringes are strongly (yet not completely) quenched
by the arrival of the second, contropolarized and Rabi-antiphases pulse.
The intensity oscillations quench can be improved upon a calibration not only of the time delay,
but also of the two pulse relative power ratio. 

\begin{figure} [ht]
   \begin{center}
   \begin{tabular}{c} 
   \includegraphics[height=5.5cm]{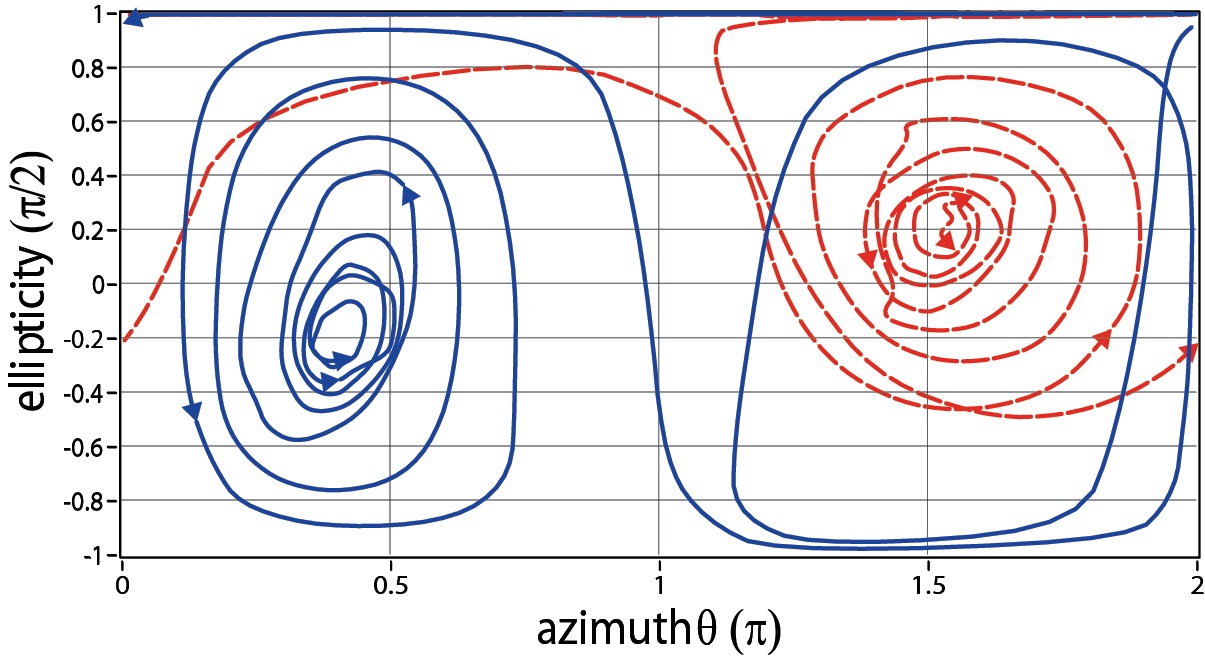}
	\end{tabular}
	\end{center}
   \caption[example] 
   { \label{fig:polar_map} 
Polarization shaping experiments plotted on the cylindrical projection of the Poincar\'{e} sphere.
The polarization oscillations are realized upon two contro-circular pulses excitation with half-integer time delay on the scale of the Rabi period: $\Phi = n\cdot 0.5~T_R$ with $n$ odd number. The final state has a degree of circular polarization (ellipticity) depending on power balance, while its specific linear direction (azimuth) is set by the optical phase delay $\phi$.}
   \end{figure}

Now,
we have already discussed
the significant lifetime imbalance between the two types of polaritons. We
know that the upper polariton enjoys a much shorter lifetime than
its homologue the lower polariton, regardless of the polarization. 
From the fits realized on the previous Rabi experiments, we have found the upper
polariton lifetime is typically of the order of 2 ps while it is of the order of 10 ps
for the lower polariton. However these short values can be now tuned by orders of
magnitudes in different samples. As an example, microcavities presenting particles
lifetime of the order of the hundred of picoseconds have been demonstrated by M.
Steger et al. (2013) \cite{steger_long_2013}. 
In our case, the lifetime imbalance between the polaritons
leads to time-dependent Rabi oscillations converging toward a monotonously
decaying signal as the population of the upper polaritons ''evaporates'' and only
lower polaritons remain.
The
decay of the polariton fields leads to a continuous drift on the Poincar\'{e} sphere,
starting by describing the initial circle of the Rabi dynamics in the absence of dissipation,
toward a final point defined by the polarized lower polaritons. This has,
as an interesting consequence, the emission of a light ''visiting'' a plethora of states
of polarization. 
In Fig.~\ref{fig:polar_map} are shown two realization cases spanning different areas of the
Poincar\'{e} sphere, which is reported as a cylindrical projextion to cartesian coordinates.
 One can see how the polarization evolves, starting from the north pole line of
right circular spin corresponding to the first right polarized pulse,
circles and starting to rotate over the sphere after the second, contro-polarized pulse arrival.
Each circle is drawn in one Rabi cycle which in the present case is about 800 fs.
The arrows on the trajectories follow the direction of time toward the final state, which here corresponds to almost linear polarization states of prevalently diagonal ($D$, on the left, solid blue trajectory) 
and anti-diagonal ($A$, on the right, dashed red trajectory) degree. 

\section{TOPOLOGY TWIST}

\lettrine{H}{ere} we show how to resonantly initialize the polariton condensate
with non-homogeneous space textures of polarization, 
and their dynamical modification in the linear regime.
In the specific, we realize a full Poincar\'{e} pattern 
 such as the skyrmion, 
whose reshaping on time lengths of the order of the polariton lifetimes
can be described as a twist in the polarization space 
and reveal a wide framework of generalized skyrmions and spin vortex states \cite{donati_twist_2016}. 
Again, the recent works we show here point
to the large potentialities of microcavity polaritons and of their topological resonant excitation
(e.g., associated to the presence of orbital angular momentum vortex states with non-trivial
polarization textures), whose dynamical redistribution represents a polarization shaping approach fully exploitable 
in engineered microcavity devices.

On the one hand, the polarization dynamics in semiconductor microcavities is
often ascribed to spin-orbital-like coupling terms, 
such as the transverse-electric--transverse-magnetic (TE-TM) modes' splitting 
resulting in the so the called optical spin Hall effect~\cite{Kavokin2005,Leyder2007,Kammann2012}.
 Such term can exert a precessing action on the pseudospin
vector, generating differently polarized regions even when starting from a homogeneous polarization~\cite{cilibrizzi_skyrmion_2016,Cilibrizzi2015}. 
However in this case, it is usually needed the presence of an imprinted or nonlinearly-activated
radial or at least inhomogeneous phase gradient, whose associated in-plane $k$ is hence differently oriented 
with respect to the initial homogeneous polarization.
On the other hand, it is sometimes useful to consider also the presence of a $k$-independent
anisotropy term $\chi_0$ associated, e.g., to strain
effects~\cite{balili_huge_2010,klopotowski_optical_2006},
that finally results in the dephasing between two given linear polarizations, 
say $x$ and $y$, and hence in the change of any non-eigenmode that can be expressed in that basis.
Such term is responsible for the polarization reshaping observed here.

The experimental realization is carried upon a modification of the previous scheme in Fig.~\ref{fig:exp_setup}.
Here we use only one of the two excitation arms on the top-left side (single pulse experiment).
However, the single excitation pulse should be thought as sculpted by two different topologies in the two spin,
respectively a $LG_{01}$ vortex of unitary winding number and a Gaussian with zero azimuthal number $LG_{00}$. 
Such configuration can be achieved by proper use of a single $q$-plate~\cite{Cardano2012},
an anisotropic and inhomogeneous liquid-crystal device
which in the last years has unfolded 
a rich exploration of optical vorticity, with recent applications to
full- and half-quantum vortex dynamics in polariton
condensates~\cite{dominici_vortex_2015} too. 
In the present case, the exciting pulse is represented by
a laser pulse (4 ps duration and 0.5 nm bandwidth) which is tuned to selectively excite only the LPB mode.
Indeed, we are not exploting the Rabi oscillations as before, 
but the peculiarity of microcavity polaritons stated above,
that consists in the anisotropy splitting of linear polarizations
along preferential axis of the samples.
By means of polarization control of the incoming/outgoing pulse
and tuning of the $q$-plate, we can set a specific superposition of two $LG$ beams with integer
or null phase winding in the two spin components:
their combination is responsible for the resultant vector vortex beam~\cite{schulz_integrated_2013}.
One relevant field pattern that is possible to obtain, among the others, is the skyrmion (see~\cite{donati_twist_2016}), 
which we imprint on the photonic pulse and hence in the initial resonant polariton state. 
Upon specific conditions the skyrmion corresponds to the stereographic projection of the whole Poincar\'{e} sphere itself
(see also \cite{beckley_full_2010}), and it should be considered the full-Poincar\'{e} beam by excellence.

   \begin{figure} [ht]
   \begin{center}
   \begin{tabular}{c} 
   \includegraphics[width=9.8cm]{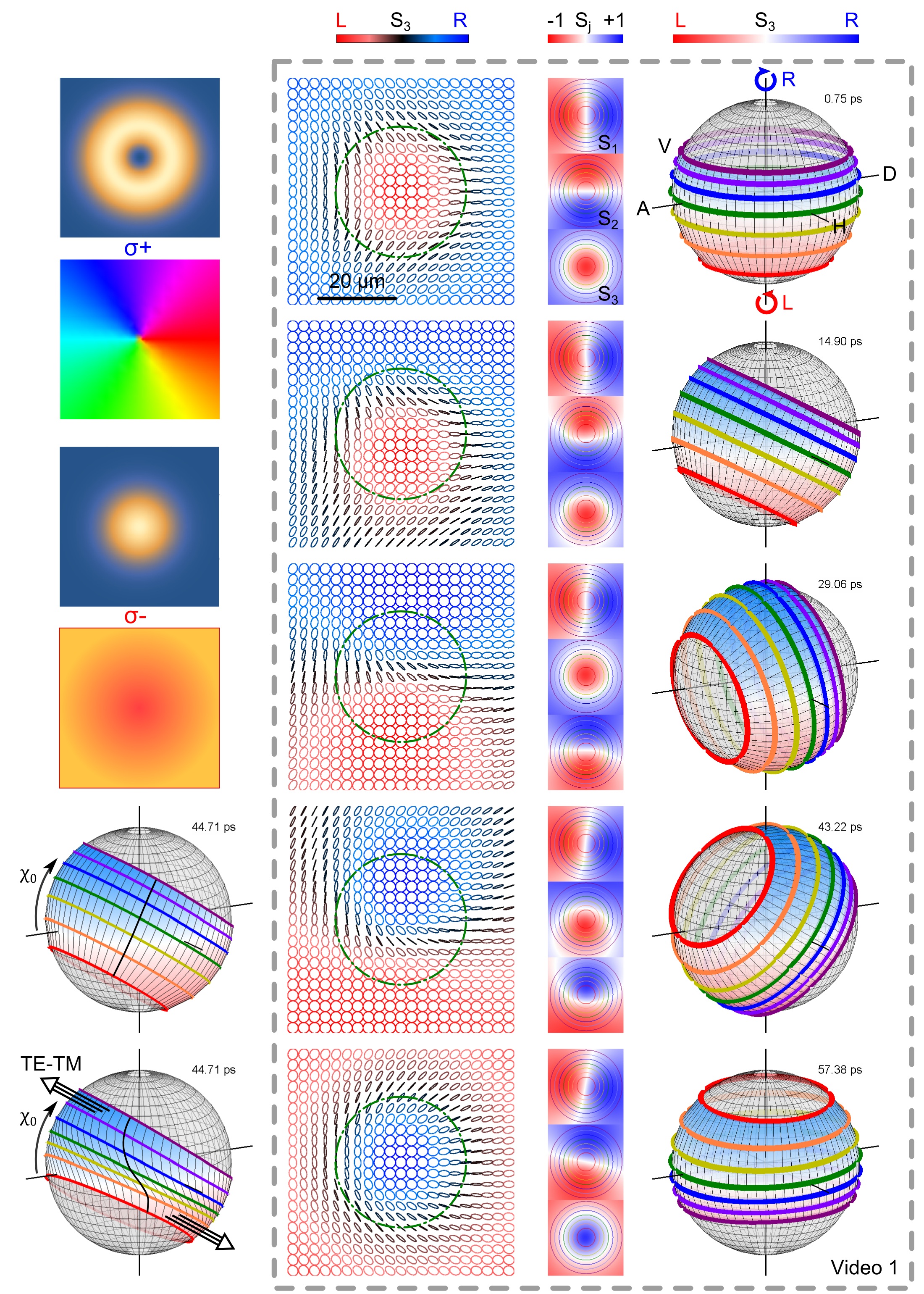}
	\end{tabular}
	\end{center}
   \caption[example] 
   { \label{fig:skyrmion_twist} 
Skyrmion twist, numerical simulation.
On the left column top panels  are reported the density and phase maps of the initial two spin components ($\sigma_+ \equiv R$ and $\sigma_- \equiv L$) of the skyrmion. The central column reports instead the polarization texture at initial and subsequent time snapshots, 
where the colour scale is proportional to the circular degree of polarization $S_3$. The $l$-line of linear polarization is reported as a dashed green circle. 
The three $S_{1,2,3}$ Stokes degree of polarizations at different times are plotted in the small panels on the right. Several concentric circles in real space (each represented with a different color) are then conformally mapped onto the Poincar\'{e} sphere on the right column. 
The anisotropy splitting $\chi_{0} = 0.08~\text{meV}$ between the horizontal ($H$) and vertical ($V$) polarizations produces a dynamical twist 
of the whole texture on the Poincar\'{e} sphere, leading at around $30~\text{ps}$ to a $90^{\circ}$ twisted skyrmion where the $S_2$ and $S_3$ distributions in real space have roughly swapped with respect to initial time. At even longer time ($\sim 60~\text{ps}$) the $180^{\circ}$ twist corresponds to a complete transmutation of the initial star skyrmion into its conjugate state (lemon skyrmion). The effect is discussed in Donati et al.~(2016)~\cite{donati_twist_2016}. The two left bottom panels ($\chi_{0} = 0.02~\text{meV}$) 
show the additional effect of including the TE-TM splitting, which leads to a transverse shrinking and an azimuthal straining of the topological belt. 
See also Supplemental Video 1.
}
   \end{figure} 

Our experimental skyrmions and their reshaping have been backed-up by theory and a numerical model,
whose results are reported in Fig.~\ref{fig:skyrmion_twist}.
As shown on the left side column,
projected onto the circular polarization basis, the skyrmions are
characterized by the presence of an integer phase winding (orbital
angular momentum) in one of the spin components and a zero-winding in
the opposite one. 
The resultant vectorial field exhibits an
inhomogeneous pattern comprising all polarization states, as typical
of full Poincar\'{e} beams (see the first panel of the central column of Fig.~\ref{fig:skyrmion_twist}). 
The skyrmions possess a specific circumference of full linear polarization ($l$-line) and they are fingerprinted by the inversion of the sign of circular polarization degree when crossing this circumference in the radial direction.
The $l$-line is conformally mapped
to the equatorial loop of the Poincar\'{e} sphere (right column) 
while the inner and outer regions to such circle
corresponds to the two hemispheres of the sphere, respectively.
Indeed the polarization changes from right-circular at the centre 
to left-circular at large distance (or viceversa).
Two fundamental types of textures can be obtained, the star-like or the lemon-like skyrmion~\cite{donati_twist_2016,bouchard_polarization_2016},
depending on the winding direction along the equatorial line with respect to that in the real-space.
In any case, the radial lines in real-space map to the meridians of the sphere,
while the circle lines in real-space map to the parallels of the sphere.
The conformal mapping of the polarization sphere to the real-space
mantains the angles between these lines.
We point out that by moving the light point of the stereographic projection into the sphere, 
the skyrmion $l$-line would gradually move towards infinitity in real-space, obtaining the polarization field of an infinite-size HQV~\cite{Rubo2007}, associated to only one emisphere.

In order to represent the dynamical behavior of our experimental observations, we
used numerical computation implementing 
 a two-component open-dissipative Gross-Pitaevskii model describing the system, 
including the anisotropy splitting term $\chi_0$.
The polarization textures reshaping at four successive time frames is reported in the central column,
highlighting a complex redistribution comprising spin transport flows across the $l$-line.
However, the reshaping is decoupled from both the mass
transport, given the linear regime, 
and from the ($\sigma_+$) phase singularity movement, which remains in the centre (not shown).
The dynamics can be projected onto each of the three Stokes basis for polarization states,
as shown in the small panels (three for each time frame) in the third column.
At the intermediate time here shown
($\sim 30~\text{ps}$),
noteworthy the $S_2$ and $S_3$ distributions in real space 
have roughly exchanged with each other, with respect to initial time.
The clearest picture of the complex reshaping appears however upon plotting on the Poincar\'{e} sphere,
as shown in the last column.
The real-space loops of the skyrmion, which initially map to the parallels on the sphere,
rotate during time around the $S_1$ basis axis associated with the anisotropy directions in real-space
(i.e., $xy \equiv HV$).
At the latest time here shown
($\sim 60~\text{ps}$) 
the twist reaches $180^{\circ}$, i.e., to the reversal of the initial star skyrmion
 into its conjugate state (lemon skyrmion).
The complete topology-flip has exchanged $R$ and $L$ spin population 
between the inner and outer regions to the $l$-line.
Similar twists are observed when starting with opposite skyrmions or double-loop topologies 
such as the hedgehog and hyperbolic spin vortices. 
 The twist speed is constant over time, as the effect is linear, independent on the
density of polaritons which are leaking out of the microcavity according to an exponential law.
The normalized twist speed retrieved from the numerical simulation
is about $\beta/\chi_0 \sim 40^{\circ}(\text{ps}\cdot\text{meV})^{-1}$,
as expected also by a theoretical evaluation.
Moreover, such velocity is the same for any circle: the real-space circular corona 
maps to a "topological belt" which is twisting rigidly.
Finally, as shown in the bottom left panels of Fig.~\ref{fig:skyrmion_twist},
 we point out that the simultaneous presence of a finite TE-TM splitting term,
is summing up to the $\chi_0$ anisotropy, resulting in a transverse shrinking and azimuthal straining 
of the topological belt.
This effect would be influenced by a large density regime,
given the nonlinear push of radial flows 
associated to finite radial $k$, summing up to the azimuthal flows of the vortex charge. 
We hope that the visualization of the reshaping dynamics of
the full Poincar\'{e} topology on the Poincar\'{e} sphere itself and the description
in terms of generalized skyrmions and spin vortices can help shedding light on
a variety of topology states and dynamics and in the perspective of possible
spintronics and polarization shaping technologies.

\section{QUANTUM VORTICES}

\lettrine{F}{ollowing} the footsteps of the previous experiments on the
coherent control of the polariton Rabi oscillations and the polariton polarization
shaping, we introduced and observed the generation of Rabi-powered
vortex oscillations in real space. This is achieved by using a series of two pulses, the first one creating a stable vortex in the polariton condensate,
and the second one disturbing the density's homogeneity leading to the
vortex oscillations. These results can be reproduced by using standard coupled-Schr\"{o}dinger Equations
 for the polaritons, fed with the adequate initial state. The
different decay sources present in the system can be added in the Hamiltonian in
order to model the oscillations damping. The system can also be solved formally
and analytical solutions can be derived. To get more insights into this dynamics,
we study the evolution of the orbital angular momentum (OAM) and see its connection
with the Rabi coupling.

Unlike their classical counterparts, quantum vortices appear to be stable configurations in a superfluid. These "topological charges", characterized by a quantized circulation $\oint \vec{v}\cdot \vec{d}l = \frac{2\pi h}{m}$ have been initially predicted by Onsager in 1949 in his work on superfluid Helium. But these predictions were only widely accepted after Feynman’s contributions in 1955. Further theoretical developments have also been made by Gross and Pitaevskii (1961). Quantum vortices have been then observed in a wide range of systems generally described by a complex wave function whose phase plays a key role in their dynamics. We can mention superconductors~\cite{blatter_vortices_1994}
where the vortex dynamics is mainly governed by the thermal and quantum fluctuations.
Such vortices can be as well manipulated by applying external forces, such as electric currents. Superconductor vortices were extensively studied as they can carry Abrikosov vortices (fluxons) and exhibit some complex ordering like Abrikosov lattices, spin glasses etc. Bose-Einstein Condensates (BEC) are also a good platform to generate quantum vortices since they are superfluid systems~\cite{leggett_superfluidity_1999}. The vortex generation in a BEC has been first demonstrated by Cornell's group in 1999~\cite{matthews_vortices_1999} following his
previous works on BEC for which he was awarded the Nobel prize in 2001. It was later demonstrated that vortices can even appear spontaneously during phase transitions in which the BEC formation occurs~\cite{weiler_spontaneous_2008}, and the reconnection of vortex lines has been studied in the framework of superfluid turbulence~\cite{Bewley2008}. Vortex lattices were also observed in rotating BEC by the group of W. Ketterle \cite{abo-shaeer_observation_2001}, exhibiting a high vortex lifetime ($\approx$ 1min). Another interesting feature of the vortices produced in atomic BECs and related to our polariton vortices behaviours, is their core's precession. The phenomenon was characterized and observed by B. P. Anderson et al. (2000) \cite{anderson_vortex_2000} in a BEC made of a superposition of two internal states of $^{87}\text{Rb}$. The vortex motion was in this case attributed to a Magnus effect applied on the vortex into the quantum fluid. More recently, it has been demonstrated that similarly to optical fields ~\cite{yao_orbital_2011,franke-arnold_advances_2008,molina-terriza_twisted_2007}, electron beams could also propagate with a phase singularity\cite{uchida_generation_2010}, giving a new
degree of freedom to the electrons. Shortly later, electron vortex beams with high
quanta of angular momentum  (up to $100\hbar$) were realized \cite{mcmorran_electron_2011,grillo_holographic_2015} leading to potential applications in electron microscopy of magnetic and biological specimens.

The polariton community was of course not left behind in the investigation
of vortex experiments. Lagoudakis et al.~(2008)~\cite{lagoudakis_quantized_2008} have reported the
formation of pinned single quantized vortices in the Bose-condensed phase of a
polariton fluid, giving new clues to understand the superfluid nature of the polariton
condensates. Quantum half-vortices were then observed in polariton 
condensates~\cite{lagoudakis_observation_2009} by the same group, following their prediction by Yu.~Rubo~\cite{rubo_half_2007}.The full-vortex twin-charge splitting~\cite{Manni2012} dynamics pinned by defects was later observed. We can also mention the observation of metastable
persistent polariton superflows~\cite{sanvitto_persistent_2010}, where it was shown how sustained
quantized angular momenta can be transferred to the steady-state condensate, insuring a long-lived vorticity. The effect of the nonlinearity on the vortex size was discussed by Krizhanovskii et al.~\cite{krizhanovskii_effect_2010} and by Voronova et al.~(2012)~\cite{Voronova2012}. More elaborated topological structures were introduced, like the formation of a vortex chain, achieved in a resonantly pumped polariton superfluid~\cite{boulier_vortex_2015}. A circular chain containing 8 vortices of the same charge was generated with a Laguerre-Gauss beam, paving the way for more experiments involving self-arranged and same-sign vortex lattices. Other recent advances on the field include the generation of unconventional states, 
such as polarization textures corresponding to spin vortices in open-cavity polaritons\cite{dufferwiel_spin_2015},
or exotic topologies (e.g., those referred to as generalized skyrmions in the previous section)
under confined nonresonant~\cite{Liu2015} and imprinted resonant~\cite{donati_twist_2016} excitation dynamics. 
Intriguing 2D+t topological strings have been associated to the half-vortex precession, 
together with pair generation phenomena at high density \cite{dominici_vortex_2015}. 
In the considered polariton system, the vortex dynamics is ruled by the interplay between the nonlinearity and the disorder landscape.

In the following, we will deal with polariton vortices in the linear regime, where the
vortices are directly imprinted to the condensate through the laser excitation. Thus,
the size of the vortices in the different components will not be affected by any nonlinearity 
as it the case for interacting condensates. In order to analyze the vortex dynamics, will also focus on the OAM value for the different polariton components. Several applications are based on the OAM properties. For example,
quantum correlations between the OAM variables can be used in quantum information~\cite{nagali_optimal_2009}s
processing and notably in protocols for quantum key distribution~\cite{leach_quantum_2010}. The OAM based multiplexing was developed and used to increase the efficiency of millimetre-wave wireless communications \cite{Yan2014} and to confer robustness to free-space communications~\cite{paterson_atmospheric_2005},
with demonstration of high accuracy transmission over distances of more than 100 km \cite{krenn_twisted_2016}. Compact plasmonic metasurfaces can be designed for the spin-orbit coupling conversion of the spin 
into an arbitrary OAM photonic vortex~\cite{bouchard_optical_2014}, while new classes of lasers have been proposed to generate all the states of an higher-order Poincar\'{e} sphere~\cite{naidoo_controlled_2016}. The data exchanged
between OAM beams was reported to reaches Terabit rates values in free-space transmission~\cite{Wang2012} and fibers~\cite{Bozinovic2013}, and recently a scheme for the time-division multiplexing of photonic pulses carrying different orbital angular momentum has been introduced~\cite{karimi_time-division_2012}. 
Topological light was also proposed to study new selection rules and manipulation of the ionized state 
in photoionization processes~\cite{picon_photoionization_2010},
thus enriching the scenarios of femtochemistry. Proposals of polariton vortices for information processing~\cite{Sigurdsson2014} have been made as well as a scheme for gyroscopes~\cite{franchetti_exploiting_2012} 
that exploits the macroscopic response of the system to small perturbations to create sensitive devices.

\subsection{Swirling Rabi vortices}

We have already said that vortex oscillations can be obtained by sending a Gaussian pulse on an unperturbed vortex, the vortex core oscillating circularly into the beam. We have experimentally observed polariton vortex oscillations that are rapidly damped after a few picoseconds, the vortex becoming immobile again. As the vortex oscillations are powered by the Rabi oscillations between the photonic and excitonic field, we suspect that the lifetime imbalance between the upper and lower polaritons plays as well a key role in this mechanism.

   \begin{figure} [ht]
   \begin{center}
   \begin{tabular}{c} 
   \includegraphics[width=11.0cm]{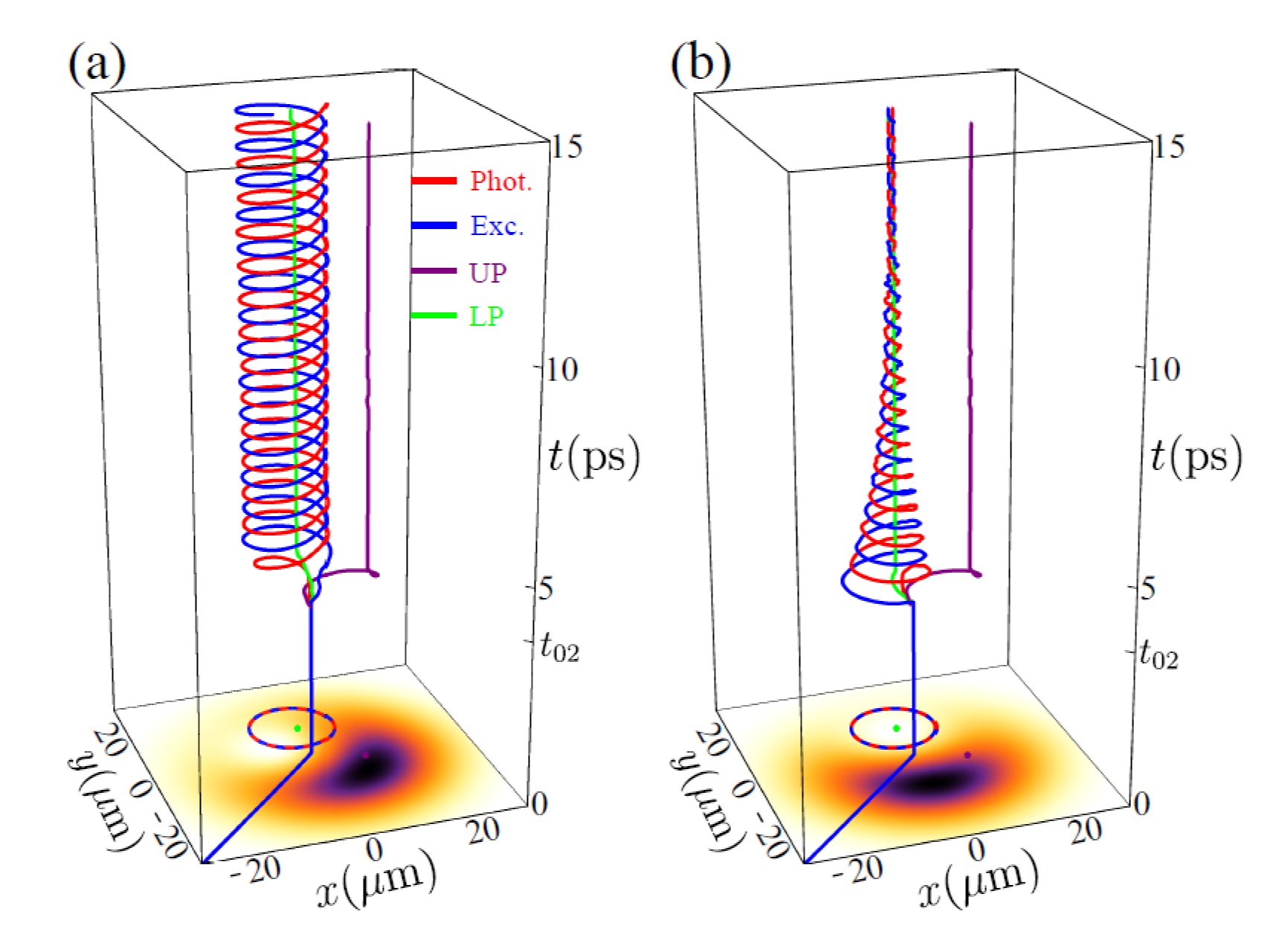}
	\end{tabular}
	\end{center}
   \caption[example] 
   { \label{fig:rartex_theo} 
Vortex oscillations generated by perturbing the system, that was initially prepared in
a vortex state, with a Gaussian pulse sent at $t = t_{02} \approx 4 \text{ps}$. The vortex trajectory in the photonic
(excitonic) field is plotted in red (blue) and also in the upper (lower) field in purple (green). The
wave function intensity $|\psi_C|$ is plotted in the $x - y$ plane for a $t > t_{02}$, where are as well projected
the vortex trajectories. (a) Conservative case, without including decay. After the second pulse, the
vortices in both fields describe circles with a non constant angular velocity. (b) Adding the upper
polariton decay, it results in a damping of the oscillations at the rate of the UP decay.
}
   \end{figure} 

The system can be simply modeled by two coupled-Schr\"{o}dinger Equations to which we add the excitation scheme, \textit{i.e.} a Laguerre-Gauss pulse (carrying the vortex phase singularity) followed by a Gaussian pulse. One can see in Fig.~\ref{fig:rartex_theo}(a) the basic vortex dynamics obtained from 
the computational model. 
Here, the Laguerre-Gauss pulse is sent at early time to generate the vortex, and the Gaussian pulse is sent at $t_{02} \approx 4 \text{ps}$. After the first pulse, the vortex is perfectly
centered into the beam and motionless. After the second pulse, the
minimum of intensity no longer remains in the center of the beam, as shown on
the density plot, which induces the vortex oscillations. The dynamical position of the
vortex is displayed with a 3D curve for both photonic (in red) and excitonic fields (in
blue). The vortices trajectory in space and time looks helical but it is actually not,
the angular velocity varying periodically in time. The two vortices, in the photon
and in the exciton, rotates around each other at the Rabi frequency similarly to a
Newton's cradle, one vortex slowing down while the other accelerates.
In the case of Fig.~\ref{fig:rartex_theo}(b) the upper polariton lifetime was included in the Hamiltonian. 
The vortex precession now follows a spiral whose radius damping corresponds to the UP lifetime, here $4 \text{ps}$. The vortex in both photonic and excitonic field converges towards the same fixed point.
To understand these vortex motions, it is interesting to look at the vortex behaviours
through the eigenstates of the system, namely by looking at the polaritonic
fields $\psi_+$ and $\psi_-$ which are obtained by combining the bare states. The vortex trajectories
in the polaritonic fields are plotted with green (lower) and purple (upper)
lines in Fig.~\ref{fig:rartex_theo}. 
Before the Gaussian pulse, the vortex is located in the center of
the beam for all the fields. After the second pulse, we see that unlike in the photon
and exciton fields, into which the vortex oscillates, it is simply shifted to the periphery
of the beam in the polaritonic fields, without exhibiting any oscillations. In the
damped case, we see now clearly that as the upper field is dying, the vortex in the
photon/exciton field converges toward the position of the lower one. We know now
that the oscillations occur because the vortex of the polaritonic fields are shifted
at different position into the beam, the vortex in the photon/exciton components
oscillating between them. Yet the vortex oscillations are located close to the vortex
position in the LP, possibly due to the mass imbalance between the two fields.

   \begin{figure} [ht]
   \begin{center}
   \begin{tabular}{c} 
   \includegraphics[width=12cm]{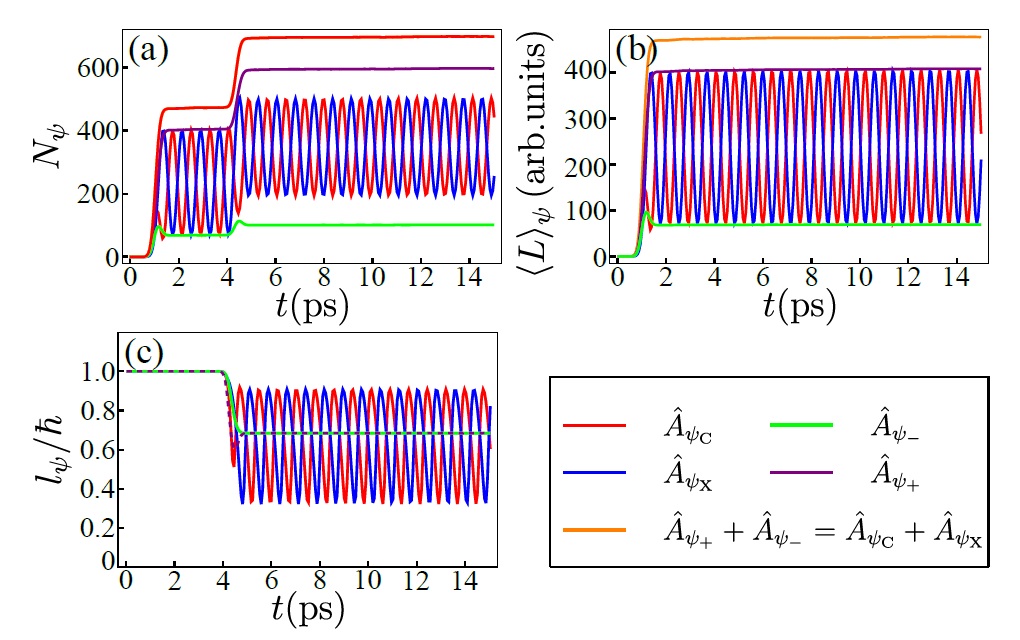}
	\end{tabular}
	\end{center}
   \caption[example] 
   { \label{fig:rartex_oam} 
Numerical dynamics of a topological two-pulse excitation.
Various OAM computed from the two-pulse experiment simulation data, corresponding
to the case of a first Gaussian pulse and a second Vortex pulse. 
(a) Fields intensities. (b) Angular Momentum for the different fields. (c)
Angular momentum per particle, normalized in units of $\hbar$. See the legend on the right bottom corner
for the color code of the operators $\widehat{A}(\langle L \rangle, N \text{~or~} l)$ applied to the corresponding field. The conserved
quantity are plotted in orange.
}
   \end{figure} 

More insight of the vortex dynamics can be obtained by looking at the value of the OAM for the different fields. For the polaritons, the OAM is conserved through the sum of the OAM of each sub-fields (photon/exciton), which allows the OAM of the individual fields to vary. In any case 
 $d(\langle \widehat{L} \rangle _{\psi_C} + \langle \widehat{L} \rangle _{\psi_X})/dt$
has to be $0$. The total number of particles $N_\psi$ is similarly conserved through the sum of the different components $N_{\psi_{\textrm{C}}}+N_{\psi_{\textrm{X}}}$. However, it is not the case for the OAM per particle defined as the ratio $l_{\psi}=\langle L \rangle_\psi / N_\psi$. In order to illustrate these conservation rules, we have computed the different OAM for the conservative case of the oscillating vortex created with two
pulses (see Fig.~\ref{fig:rartex_theo}(a)). One can see in Fig.~\ref{fig:rartex_oam}(a) the time evolution of the
fields intensities $N_{\psi}$. The bare fields  $\psi_C$ (in red) and $\psi_X$ (in blue) exhibit Rabi oscillations while the polaritonic fields  $\psi_+$ (in purple) and $\psi_-$ (in green) do not. The different intensities of course increase after the second pulse arrival, the system being filled with more particles. The
conserved quantities are plotted in orange. The OAM is plotted in (b) for the different fields. One can see that the OAM of the bare fields oscillates exactly between the OAM’s steady value of the polaritonic fields.
The OAM for the different fields is not be affected by the second pulse, that initiates the vortex oscillations process. Which makes sense considering that the Gaussian does not carry additional angular momentum.
Nevertheless, the vortex oscillations appear more explicitly when one computes the OAM per particle for the different fields, as plotted in (c). When the vortices are exactly in the center of the beam, that is before the second pulse,
the angular momentum per particle is exactly quantized to one unit of $\hbar$.  After the second pulse, the vortices in the polariton fields do not oscillate but are displaced to a fixed position on the periphery of the beam (see purple and green lines in Fig.~\ref{fig:rartex_theo}), with a corresponding value of $l$ which is lower than 1. As mentioned by Pitaevskii and Stringari~\cite{pitaevskii_book03a}, for a vortex core displaced from the center of the beam, the OAM per particle takes a value lower than $1$, the axial symmetry of the problem being lost. Here, the
polaritonic vortices are symmetrically displaced into the beam, $l_{\psi_+}$ and $l_{\psi_-}$ thus share the same value. In the photonic and excitonic fields, the vortices oscillate from the center to the periphery in circle, so that the variation range of the OAM per particle oscillations is constant. Since the vortex motion is directly powered by the Rabi oscillations, we expect that the accurate control of these new topological strings could be achieved in the same way that we performed the control of the quantum state or the polarization state in our previous experiments.

\acknowledgments 
 
We acknowledge Romuald Houdr\'{e} for the growth of the microcavity polariton device and Alberto Bramati for the related know-how, Lorenzo Marrucci and Bruno Piccirillo for the $q$-plate devices and the related know-how. 
We acknowledge the project ERC POLAFLOW (Grant 308136) for
financial support.


\end{document}